\documentclass[journal,11pt,onecolumn,draftclsnofoot,]{IEEEtran}

\usepackage{xspace}
\ifCLASSOPTIONcompsoc
  \usepackage[nocompress]{cite}
\else
  \usepackage{cite}
\fi
\ifCLASSINFOpdf
   \usepackage[pdftex]{graphicx}
\else
\fi

\usepackage{multirow} % Required for multirows
\usepackage{booktabs} % For prettier table

\newcommand{\ie}{\emph{i.e.,}\xspace}
\newcommand{\eg}{\emph{e.g.,}\xspace}
\newcommand{\etal}{\emph{et al.}\xspace}

\newcommand{\coli}{\emph{E. coli}\xspace}
\newcommand{\yeast}{\emph{S. cerevisiae}\xspace}
\newcommand{\fly}{\emph{D. melanogaster}\xspace}
\newcommand{\human}{\emph{H. sapiens}\xspace}
\newcommand{\gcnlayer}[2]{\ensuremath{\vec{h}^{(#1)}_{#2}}}
\newcommand{\weights}[1]{\ensuremath{\mathbf{W}^{(#1)}}}

\begin{document}
%\title{Gene Essentiality Transduction With Graph Neural Networks}
\title{EPGAT: Gene Essentiality Prediction With\\ Graph Attention Networks}

\author{João~Schapke,~Anderson~Tavares,~and~Mariana~Recamonde-Mendoza% <-this % stops a space
\IEEEcompsocitemizethanks{\IEEEcompsocthanksitem J. Schapke, A. Tavares, and M. Recamonde-Mendoza are with the Institute of Informatics, Universidade Federal do Rio Grande do Sul, Porto Alegre, Brazil. E-mail: joaoschapke@gmail.com, \{artavares, mrmendoza\}@inf.ufrgs.br.}% <-this % stops an unwanted space
\thanks{Preprint. This work has been submitted to the IEEE for possible publication. Copyright may be transferred without notice, after which this version may no longer be accessible.}}


\markboth{}%
{Schapke \MakeLowercase{\textit{et al.}}: TITLE}



\IEEEtitleabstractindextext{%
\begin{abstract}

The identification of essential genes/proteins is a critical step towards a better understanding of human biology and pathology. Computational approaches helped to mitigate experimental constraints by exploring machine learning (ML) methods and the correlation of essentiality with biological information, especially protein-protein interaction (PPI) networks, to predict essential genes. Nonetheless, their performance is still limited, as network-based centralities are not exclusive proxies of essentiality, and traditional ML methods are unable to learn from non-Euclidean domains such as graphs. Given these limitations, we proposed EPGAT, an approach for essentiality prediction based on Graph Attention Networks (GATs), which are attention-based Graph Neural Networks (GNNs) that operate on graph-structured data. Our model directly learns patterns of gene essentiality from PPI networks, integrating additional evidence from multiomics data encoded as node attributes. We benchmarked EPGAT for four organisms, including humans, accurately predicting gene essentiality with AUC score ranging from 0.78 to 0.97. Our model significantly outperformed network-based and shallow ML-based methods and achieved a very competitive performance against the state-of-the-art node2vec embedding method. Notably, EPGAT was the most robust approach in scenarios with limited and imbalanced training data. Thus, the proposed approach offers a powerful and effective way to identify essential genes and proteins.

%The survival of living organisms is dependent of only a subset of genes within it's genome, coined essential genes. The study and discovery of these essential genes is of immeasurable value as they give us understanding of life and the intricate relations of the genome. Essentiality can be estimated from information on the interactions among genes within the genome. Gene interaction is implicitly encoded in protein-protein interaction (PPI) datasets, which is the reason why current methods for gene essentiality prediction rely mostly on PPI networks. But, current methods estimate essentiality by measuring correlated information to essentiality (e.g. the amount of interactions a gene has), which are only proxies for the real value being estimated. In this article we aim to directly derive predictions from PPI networks using Deep Learning models, specifically Graph Neural Networks (GNN), which directly map raw graph data to estimated properties. We benchmark GNNs and current approaches on a set of PPI networks of organisms with known essentialome and show that GNNs provide improved predictive performance. 
\end{abstract}

}

\maketitle

\IEEEdisplaynontitleabstractindextext

% \fi
\IEEEpeerreviewmaketitle

\section{Introduction}\label{sec:introduction}


%contexto / motivacao - pq o problema eh interessante?
\IEEEPARstart{D}{iscovering} the essentialome, \ie the set of essential genes for the survival of a living organism, is an important research question in bioinformatics and genomics. A plethora of fundamental functions are played in the cell by proteins coded by essential genes \cite{essenceOfLife}, resulting in an irreplaceable role in controlling cellular processes. Therefore, essential genes are regarded as the basis of life, such that characterizing an organism's essentialome can give us insights into the inner working and interdependence within its genome.  Knowledge of essential proteins and genes not only improves our understanding about biological processes and molecular functions, but also has contributed to a range of fields, such as in synthetic biology, for the definition of a minimal genome  \cite{minimalGenes}; in drug design, for the choice of drug targets of antimicrobial and anticancer compounds  \cite{cancer,drugtarget}; in a better understanding of diseases, by helping discover pathogenic genes \cite{park2008analysis}; and in metabolic engineering \cite{metabolic}.

Several experimental procedures, such as targeted single-gene knockout  %\cite{experimentalKnockout1,experimentalKnockout2}
and transposon mutagenesis, %\cite{experimentalMutagenesis}, 
have been developed to identify essential genes \cite{rancati2018emerging}. Although they are generally robust and accurate, wet-lab approaches imply costs with laboratory resources and are time consuming, which may impose significant limitations to their use, especially for large scale analyses and for complex organisms like mice and humans \cite{survey}. 
%In addition, experimental techniques hardly apply to some complex organisms like mice and humans \cite{survey}. 
Considering this scenario, computational approaches have become appealing solutions as they provide fast and inexpensive results, which can reduce the workload in experimental procedures by suggesting a list of candidate genes and proteins to be tested. With the rapid development of high-throughput techniques, a large volume of biological data is available for \textit{in silico} studies, covering a wide variety of organisms. This factor was crucial for the further development of computational methods to predict essential genes and proteins.

%desafios - pq o problema eh dificil?
As recently reviewed \cite{survey}, computational methods usually rely on network topology features extracted from Protein-Protein Interaction (PPI) networks as the main source of data to predict gene essentiality. These methods leverage the fact that highly connected genes\footnote{We note that although PPI networks encode interactions among proteins, we refer to the nodes of the network as genes or proteins interchangeably throughout the text, considering the genes by which these proteins are encoded.} in the PPI network are more likely to be essential -- the \textit{centrality-lethality} rule \cite{whyhubs}. Therefore, correlation between node centralities and gene essentiality has been explored in several computational approaches previously proposed, including neighborhood-based methods such as degree centrality (DC)\cite{Jeong2001}, local average connectivity (LAC)\cite{lac}, and edge-clustering coefficient centrality (NC) \cite{nc}.

%This fact led to a plethora of essentiality prediction methods based on network theory, \eg \cite{nc,lac,Jeong2001}. Thus, most methods developed so far rely on analytical measures of network properties or simple message passing propagation between genes. 

Although information regarding interaction among genes is strongly related to gene essentiality, using only PPI network topological features leads to limited prediction accuracy \cite{zhang2016review}. %do bring some caveats.
First, PPI networks are known for being noisy and containing many false positive connections \cite{fp}. Second, there are many protein interaction databases, with significantly different networks among them, which not only emphasizes the potential incompleteness of single PPI networks, but may also generate varying and unstable results depending on the network used. Third, due to the complexity of living organisms, it is natural to assume that gene essentiality is determined by multiple biological factors that can not be fully captured by topological characteristics of the network \cite{rancati2018emerging}.

%Soluções atuais e suas limitações
Recent works aim at alleviating these issues by using multiomics datasets to either manually prune untrustworthy connections or as additional input for the statistical and computational methods being applied. %The idea behind data integration is to provide a more robust and holistic view of the system, thus contributing to the extraction of better hypotheses on gene essentiality. 
In this direction, gene expression profiles are perhaps the most common choice of additional evidence among works in the field \cite{survey}, being applied, for instance, in \cite{ref4} and \cite{ref6}. % and showed to improve the performance of models in our experiments significantly, 
Other types of biological information already used in conjunction with PPI networks are subcellular localization information \cite{peng2007rechecking}, orthology information \cite{li2018united}, protein complexes \cite{li2015united} and domains \cite{peng2014udonc}, sequence data \cite{deng2011investigating}, and functional annotations such as Gene Ontology\cite{kim2012prediction}.
In addition, there is an increasing number of studies exploring three or more sources of biological information in the construction of statistical or computational models (\eg \cite{li2016reliable,li2016predicting,lei2018predicting}).

%Also relating to our work are other approaches that have explored integrating multiomics data. The use of multiomics datasets it's an important development within bioinformatics and in our task as well. We referred ourselves to previous work for our choice of omics data. Gene expression profiles are a common choice among work on the field \cite{ref4} \cite{ref6} and showed to improve the performance of models in our experiments significantly, subcellular localization information was previously adopted by \cite{ref5} and \cite{ref7}, and orthology information by \cite{ref5}. Other types of data not explored in our work are protein complexes \cite{ref8} and GO annotations \cite{ref10}.

Given the nature of this prediction problem, computational approaches often apply machine learning (ML) algorithms, especially supervised learning methods, as part of the solution  \cite{zhang2016review}. In this sense, the identification of essential genes and proteins is regarded as a binary classification task, and algorithms build a prediction model based on features related to gene and protein essentiality. Most works in this scope have applied shallow ML algorithms, among which Support Vector Machines (SVM)\cite{svm}, %random forest (RF), %ensemble-based, 
na\"ive Bayes \cite{naivebayes}, and decision trees \cite{tree} are recurrent ones. This methodology undoubtedly helped advance the frontier of knowledge regarding essential genes and proteins. Nonetheless, whereas centrality-based methods needs prior knowledge to create good score functions, shallow ML methods demand the specialist's input to select representative biological properties as features for the learning problem, which is not a trivial task, especially when knowledge regarding these features is still evolving and may not be fully elucidated % and under consolidation 
\cite{rancati2018emerging}. 

Deep learning methods, on the contrary, avoid the critical step of extracting handcrafted features in the design of ML methods, automatically learning features from the training data during its operation.
%since this step is embedded in the algorithm's own functioning. Deep learning automatically learns features from the training data,   
Thus, deep learning does not require prior knowledge or assumptions regarding relevant biological features of gene essentiality. Moreover, deep learning methods have demonstrated an excellent performance in a wide range of prediction and network-based problems in Bioinformatics \cite{min2017deep,jin2020application}.

Zeng \etal \cite{zeng2019deep} proposed a deep learning
framework for identifying essential proteins (DeepEP) without
any prior knowledge, adopting %three different types of biological information, namely 
PPI network, gene expression data, and subcellular localization information. The node2vec algorithm \cite{grover2016node2vec} was applied for network representation learning, a long short term memory network was used to process the gene expression data, and the output vectors originated from pre-processing the three data sources are concatenated and classified by a fully connected layer with the sigmoid activation function. 
%unsupervised feature learning to 
%automatically extracting low-dimensional semantic and topological features from the network. 
Computational experiments with the PPI
network of \emph{Saccharomyces cerevisiae} indicate
an area under the Receiver Operating Characteristic (ROC) curve (AUC) of 0.832.

Also in this direction, PPI networks and gene expression data were used as input for the deep learning framework proposed in \cite{deepep}. The node2vec technique was used for encoding the PPI network into a low-dimensional space, which is concatenated with patterns extracted by a multi-scale Convolutional Neural Network (CNN) from an image-based representation of gene expression profiles. The output vector is analyzed by a sequence of two fully connected layers using the rectified linear unit (ReLU) and softmax activation functions to predict the final label of a protein. An AUC score of 0.82 is achieved for \yeast data. The works by \cite{zeng2019deep} and \cite{deepep} show that the dense vectors generated by node2vec technique contribute to an improved performance over commonly used centrality measures.

More recently, Zhang \etal \cite{deephe} proposed DeepHE, a deep learning-based method to predict human essential genes. DeepHE integrates sequence features extracted from DNA and protein sequences with features learned from
PPI network using node2vec. A deep neural network was trained with a cost-sensitive technique to address the imbalanced learning problem inherent to this domain, predicting human gene essentiality with an average AUC score higher than 0.94. The method is shown to outperform traditional shallow ML algorithms such as SVM, RF, and na\"ive Bayes.

Despite the success of these deep learning methods for gene or protein essentiality prediction, they have in common the need to use a node embedding technique (\eg node2vec) to transform the PPI network to an ordered and fixed-size input lying in the Euclidean space, as expected by statistical and ML models. In other words, these approaches do not learn from the full graph data, but rather from low-dimensional representations obtained from them. Therefore, it is reasonable to assume that during this transformation, valuable information may be lost and more complex patterns may not be discovered, impacting in the performance and generalization power of the model.

Graph Neural Networks (GNNs) \cite{gori2005new,scarselli2008graph} were introduced as a class of deep learning based methods capable of dealing with the non-Euclidean nature of graphs, automatically learning network topology-preserving node-level vector representations from networks. %learning from graph-structured data using specially designed neural layers. 
%GNNs discard the need for encoding node's information, as they can do propagation guided by the graph structure instead of using it as part of features.
Graph Convolutional Networks (GCNs) \cite{kipf2016semi} are the current state-of-the-art for problems posed for GNNs, performing node classification based on node features and network topology by aggregating information from neighboring nodes in a hierarchical fashion. GCNs have significantly improved performance in other prediction tasks in Bioinformatics, such as estimating drug–target binding affinity \cite{nguyen2019graphdta}, identifying cancer driver genes \cite{schulte2019graph}, and classifying breast cancer subtype  \cite{rhee2017hybrid}.

In this work, we hypothesize that GNN-based models could offer high performance in the identification of essential genes and proteins compared to previous deep classifiers, by learning more complex relations directly from the PPI networks. We propose EPGAT, a novel computational method for gene essentiality prediction based on Graph Attention Networks (GATs) \cite{gat}, an extension to GCNs that adds an attention mechanism to the original model. As far as we are aware, no previous work has adopted GCNs, or more especifically GATs, for prediction of essential genes and proteins.

Our approach aims at tackling the main limitations of current methods, \ie (i) the low reliability of PPI networks, (ii) the need to integrate multiomics data to more broadly capture the biological notion of essentiality, and (iii) the limited use of graph-embedded knowledge by previously proposed network topology measures, shallow ML algorithms, and deep learning methods. %the use of graph-like structure by prediction methods. 
%, by proposing a novel solution based on Graph Convolutional Networks, which have not been explored in this domain.
Our solution operates directly over the graph structure by using GATs, and incorporates multiomics datasets, namely gene expression profiles, orthology information, and subcelular localization information, as node features, which are collectively used with PPI networks in the model learning process. %and follow the latter approach. 
To deal with issue (i), we evaluate our experiments on three distinct PPI network databases. Additionally, we benchmark our method on four organisms: \emph{Escherichia coli}, \emph{Saccharomyces cerevisiae}, \emph{Drosophila melanogaster}, and \emph{Homo sapiens}.  %Our experiments sho show that GATs provide a simpler and better performing approach for prediction of gene and protein essentiality. 
Our contributions in this article are: 1) We show that EPGAT provides state-of-the-art results compared to other ML and network-based approaches and 2) we analyze and integrate different PPI and multiomics datasets in order to infer which of them are the most relevant for gene essentiality prediction.

% explicar a estrutura do texto.
The remainder of this paper is organized as follows. We introduce the relevant background and the setting of our experiments throughout Section~\ref{S:Materials&Methods}. In Section~\ref{S:Experiments&Results}, we present our experiments, results, and discussion of our findings. %, followed by a discussion of our findings in Section~\ref{S:Discussion}. 
Lastly, in Section~\ref{sec:conclusion}, we conclude our study and offer perspectives for future research.

\section{Materials and Methods}
In this section, we present the basis of our proposed method, including an overview of the used data and learning algorithm, evaluation strategies, and baselines adopted for comparison purpose.

\label{method}
\label{S:Materials&Methods}

\subsection{Data Collection and Preprocessing}
\label{S:Data}
The proposed method was evaluated in the prediction of essential genes on four organisms: \emph{Escherichia coli}, \emph{Saccharomyces cerevisiae}, \emph{Drosophila melanogaster}, and \emph{Homo sapiens}. 
\yeast and  \coli  were choosen as they have their complete essentialome published \cite{yeastEssential}\cite{coliEssential} and are widely used as testbed for statistical models. 
We also tested our method on the human genome since the results and performance of computational models in complex organisms that do not have a known essentialome is an open question and, therefore, a valuable research direction. %a relevant indicative of its usefulness. % , such as humans, is more relevant. 
\fly is also an interesting benchmark organism because, as we will later discuss, it is the most negatively biased organism of our dataset regarding annotation on essential genes.
%as it is the most negatively biased organism, out of our dataset of 12.000 labeled genes only 200 are essential. 

Besides essential genes dataset, we used four kinds of biological datasets in our experiments: PPI networks, gene expression profiles, subcellular localization, and orthology information. To standardize nomenclature among these sources and allow data integration, the identification of genes and proteins was mapped to UniProtKB nomenclature. Elements without such correspondent were removed from our dataset. In what follows we explain our data sources.

\subsubsection{Essential Genes Dataset} 
Annotations for essential and non-essential genes for all organisms 
%The labels for essential genes of all organisms 
were downloaded from the database of Online GEne Essentiality (OGEE) (downloaded at 25/03/2020) \cite{ogee}.  
A total of 5,636 genes (18.63\% essential) were obtained for \yeast,  
4,322 genes (8.2\% essential) for \coli, 
13,781 genes (2.96\% essential) for \fly, 
and 21,556 (8.9\% essential) genes for \human.

After the pre-processing step to standardize the nomenclature of genes using the UniProKB standard, the number of genes labeled for each species was: 5,636 genes (18.63\% essential) for \yeast,  
3,686 genes (6.56\% essential) for \coli, 
12,329 genes (1.95\% essential) for \fly, 
and 18,476 genes  (9.88\% essential) for \human.  As we may observe, \fly is the most negatively biased organism since only 1.95\% of labeled data refer to essential genes (\ie positive instances, in the context of classification).%out of our dataset of 12,329 labeled genes only 200 are essential.

\begin{table}[!t]
  \begin{center}
    \caption{Characteristics of the PPI networks used in this study.}
     \label{datasets_table}
    \resizebox{0.8\columnwidth}{!}{%
    \begin{tabular}{l|c|c|c|c|c|c} 
    \toprule % <-- Toprule here
     {} &  \multicolumn{3}{c|}{\textbf{\yeast}} & \multicolumn{3}{c}{\textbf{\coli}} \\
    \midrule % <-- Toprule here
     {}  & \textbf{BioGRID} & \textbf{STRING} & \textbf{DIP} & \textbf{BioGRID} & \textbf{STRING} & \textbf{DIP} \\
      \midrule % <-- Midrule here
            N. nodes & 6908 & 6049 & 5126 & 3971 & 4068 & 2924 \\
            N. edges & 526392 & 393022 & 22941 & 178271 & 114432 & 12246 \\
            N. labeled genes & 5548 & 5455 & 4570  & 3489 & 3521 & 1902 \\
            N. essential genes & 1048 & 1050 & 981 & 201 & 234 & 205 \\
            N. test labels & 1110 & 1091 & 914 & 698 & 705 & 381 \\
      \midrule % <-- Midrule here
           \multirow{2}{*}{} &  \multicolumn{3}{c|}{\textbf{\human}} & \multicolumn{3}{c}{\textbf{\fly}} \\
      \midrule % <-- Midrule here 
      {}  & \textbf{BioGRID} & \textbf{STRING} & \textbf{DIP} & \textbf{BioGRID} & \textbf{STRING} & \textbf{DIP} \\
       \midrule % <-- Midrule here 
            N. nodes & 16562 & 18822 & 4615 & 2484 & 11499 &  972 \\
            N. edges &  479496 & 1340788 & 7417 & 22653& 622980 & 1401 \\
            N. labeled genes & 14486 & 17894 & 3370 & 1778 & 8592 & 625 \\
            N. essential genes & 1590 & 1806 & 726 & 79 & 161 & 41 \\
            N. test labels & 2898 & 3579 & 674 & 356 & 1719 & 125 \\
      \bottomrule % <-- Bottomrule here
    \end{tabular}
    }
  \end{center}
\end{table}

\subsubsection{PPI Networks}
%PPI networks contain relevant information for the prediction of essential genes and were used as the graph data for our GAT-based prediction method.
As previously reported, PPI networks have many false positive interactions\cite{fp}, and networks for the same organism from different databases are extremely heterogeneous. 
This may lead to results instability due to the choice of PPI dataset. 
To analyze and alleviate the effect of network choice, we evaluated the results for each organism in three different PPI networks: DIP (data as of 2017-02-05), BioGRID (Version 3.5.182), and STRING (Version 11.0).

\textbf{DIP}, the Database of Interacting Proteins \cite{dip}, lists protein pairs that were experimentally shown to bind to each other. %Multiple sources of data are analyzed and combined, and data curation is performed both manually by experts and by automatic computational approaches. 
Although DIP has been used by previous works, it has the characteristic of being a sparse network. In our work, DIP dataset contains the lowest number of nodes and edges among all networks used, for all organisms evaluated.

\textbf{BioGRID}, the Biological General Repository for Interaction Datasets \cite{biogrid}, provides the curation and storage of protein interactions reported in the biomedical literature for all major model organisms and for humans. Interactions in this database are divided between physical and genetic, both of which are used in our dataset.  %To date, BioGRID has over 1.6 million pairs of interactions catalogued. 
%BioGRID dataset is a much denser network in comparison to DIP. 

\textbf{STRING} \cite{string} dataset assembles PPI collected and scored from different %interactions from different sources
'evidence channels', depending on the origin and type of the biological evidence. %There are seven channels, including \textit{experiments}, \textit{text mining}, and \textit{database}. 
A combined and final score is computed for each interaction, which is typically used as an estimate of the likelihood that a given interaction is biologically meaningful, given the supporting evidence. Heuristically, we filtered out connections with a confidence score bellow 0.5 aiming at reducing false positive interactions. As part of our experimental approach, we further analyzed such choice, as discussed in Section~\ref{S:PPIThreshold}. 

DIP dataset lists only proteins that bind to each other, while STRING and BioGRID may also include indirect interactions between proteins, which makes them much denser. We also note that we added self-connections for every node in the networks, a standard procedure in order to train GATs and GCNs in general (see Section~\ref{S:GNNs} for further details).

Details of collected PPI networks are summarized in Table~\ref{datasets_table}. The number of nodes and edges refer to the structure of networks collected from the databases aforementioned. 
The number of labeled genes of each PPI network is the number of genes contained within the essential genes dataset that intersect with elements from the PPI network, including both positive and negative labels regarding essentiality. The positive labels, which are of special interest in our work, are shown in the "N. essential genes" field. 
%The number of true essential genes (\ie positive labels in the essential genes dataset) in this intersection is shown as the number of positive labels. 
%The number of test labels is the partition of genes within the intersection reserved for evaluating the models using random train-test splits.
Finally, the number of test labels is the size of the partitions reserved for model evaluation from the labeled genes dataset. %within the intersection reserved for evaluating the models.% in each random train-test split.

\subsubsection{Gene Expression Profiles}

In our multiomics-based approach, we used gene expression profiles from the Gene Expression Omnibus (GEO) database\cite{geo}.
For \coli, we collected gene expression data from \textit{GSE7326} and \textit{GSE40693}, which we found empirically to be highly correlated with the \coli essentialome. \textit{GSE7326} evaluates \coli gene expression during cell death, whereas \textit{GSE40693} measures the transcriptomic changes of \coli in response to an antimicrobial compound. For \yeast, we used the expression profiles from the \textit{GSE3431} dataset, which was previously applied for essential gene prediction \cite{ref3}. %expression profile
\textit{GSE86354} and \textit{GSE67547} were used for \human and \fly, respectively. \textit{GSE86354} provides expression profiles for 1,558 samples across 8 tissue sites generated by the Genotype-Tissue Expression (GTEx) project, and \textit{GSE67547} expression is obtained over the lifespan of 120,000 fruit-flies. Further information about gene expression data may be obtained in the GEO database through datasets' accession numbers.

\subsubsection{Subcelullar Localization and Orthology Information}
As additional biological evidence, we used gene orthology information gathered from the \textit{InParanoid} (v8) \cite{inparanoid} database and protein subcellular localization data from the COMPARTMENTS database (version 8) \cite{compartments}. We note that the COMPARTMENTS databases does not provide information on subcellular localization for \coli. For this reason, we evaluate this organism using only the gene expression and orthology information as additional data.

\subsection{Graph Neural Networks}
\label{S:GNNs}

%A graph $\mathcal{G}$ is denoted by a set of nodes $\mathcal{N} = \{1, \ldots, N\}$, representing entities, and a set of edges $\mathcal{E}$, which connect these entities.
%As an example, in the graph of a protein-protein interaction (PPI) network, nodes are proteins and edges represent interactions among these proteins based on some biological evidence. 

%In a prediction problem, traditional ML methods usually rely on pre-processed data, where features are extracted from instances to generate a tabular dataset.
%As neighborhood information is important in a graph, the entry of each node in the resulting tabular dataset would need to contain features of neighboring nodes. 
%As the number of neighbors for each node varies, the resulting dataset would be unwieldy.
%Hence, methods that deal with raw data are desirable.

A PPI dataset may be denoted by a graph $\mathcal{G}$ composed by a set of nodes $\mathcal{N} = \{1, \ldots, N\}$, representing proteins, and a set of edges $\mathcal{E}$ reflecting the interactions among the proteins.
Additionally, we may represent information of each protein in the network by adding attributes to the node it is represented by. 
With this reasoning we frame essential gene prediction as a node classification problem over a PPI network attributed with additional multiomics data.

Traditional ML methods usually rely on preprocessing procedures in order to parse graph data and generate a tabular dataset.
As neighborhood information is important in a graph, the entry of each node in the resulting tabular dataset would need to contain features of neighboring nodes. 
As the number of neighbors for each node varies, the resulting dataset would be unwieldy.
Hence, methods that deal with raw graph data are desirable.

Graph Neural Networks (GNNs) comprise a family of ML methods that handle graph data without pre-processing. 
In this work we use a specific GNN method, namely Graph Attention Networks (GATs) \cite{gat}. GATs combine ideas of generalized convolutions  \cite{kipf2016semi}, which allows graph nodes to aggregate information from their irregular neighborhoods, with self-attention mechanisms \cite{Vaswani2017attention}, which allows nodes to learn the relative importance of each neighbor during the aggregation process. 
Next, we describe GATs for node classification tasks, where the goal is to predict the class $c$ of each graph node out of $C$ possible classes.

In a GAT, each layer $l \in \{1,  \ldots, L\}$ contains a vector-valued representation, also called embedding, \gcnlayer{l}{i}, for each node  $i \in \mathcal{N}$.
For an arbitrary node $i$, the first layer, \gcnlayer{1}{i} may contain features that reflect properties of the node (e.g. multiomics protein data), domain knowledge or just be randomly initialized. 
Embeddings of intermediate (hidden) layers represent increasingly ``higher-level'' latent features of the nodes. 
Embedding dimensions can be arbitrary in the hidden layers.
The final layer,  \gcnlayer{L}{i}, is a vector of dimension $C$, containing the probability of node $i$ to belong to each one of the $C$ classes.

For a given node $i$, the embeddings of all its neighbors $j$ in a layer, \gcnlayer{l}{j}, are used to compute $i$'s embedding on the next layer's \gcnlayer{l+1}{i}. 
To make use of its own embedding, the set of $i$'s neighbors, $\mathcal{N}_i$, includes $i$ itself.
Equation \ref{eq:gcn} shows the embedding calculation procedure, where $\sigma$ is a non-linear activation function, applied element-wise to the resulting vector,  \weights{l} is the matrix of learnable weights of layer $l$ and $\alpha_{ij} \in [0,1]$ is the learnable attention between $i$ and $j$, which indicates how strongly $i$ ``listens'' to information on node $j$.

\begin{equation}
\gcnlayer{l+1}{i} = \sigma( \sum_{j \in \mathcal{N}_i} \alpha_{ij} \weights{l} \gcnlayer{l}{j})
\label{eq:gcn}
\end{equation}

Thus, the goal in GATs is to adjust the weights $\weights{l}$ of each layer and the attention coefficients $\alpha_{ij}$ between every pair $i,j$ of nodes so as to capture latent node features and, ultimately, generate good class predictions.
Note that the weights of each layer, $\weights{l}$, are shared, \ie all nodes use the same weight matrix to update their next layer's embeddings. 
Hence the importance of the self-attention mechanism: it weighs the importance of each neighbor when updating a node's embedding. 
Moreover, this value is also learned from data.

As node embeddings lie in a continuous space, %and in allow choosing differentiable activation functions
GATs can be seen as a deep learning technique. 
This means that, if differentiable activation functions are used, GATs can be trained end-to-end with gradient descent methods.

\subsection{Model Training and Performance Evaluation}
\label{S:Training}

For model training and evaluation, we randomly split the labeled data into a training (80\%) and a test set (20\%) in a stratified manner, \ie preserving original classes proportions in both sets. To better estimate the models generalization power, ten random splits were used and performance metrics were averaged across repetitions. 
Additionally, we reserved a subset of the trainining dataset and used it for validation while training statistical methods (including GAT). The validation data is used for early stoppage of the training procedure.

Essential genes compose only a small fraction of the coding genome of most organisms, including \coli, \yeast, \fly, and \human. As a result, datasets of essential genes are heavily inclined towards negative cases. To account for this class imbalance, we trained EPGAT using the weighted binary cross-entropy (CE) function, defined as:

\begin{equation}
CE = -c_1 y log(p) - c_0 (1-y) log(1-p)
\label{eq:crossentropy}
\end{equation}
where $y$ is the true label of the instance and assumed to be 1 (\ie positive or essential) or 0 (\ie negative or non-essential), $p$ is the probability predicted by the model for the positive class (and thus $1-p$ is the probability predicted for the negative class), and $c_1$ and $c_0$ are inversely proportional to the number of instances  in the training dataset for the positive and negative classes, respectively.

Training GAT models involves tuning several hyperparameters, which can be very time consuming, especially when a number of models are being developed, as in our work. To reduce computational costs, we first determined the best combination of hyperparameters for the \yeast networks using optuna \cite{akiba2019optuna}, an algorithm for hyperparameter optimization in ML. Next, based on the results from this analysis, we performed an empirical evaluation for the other organisms by manually varying the hyperparameters using a smaller number of combinations that had given good results for the \yeast networks. 
The hyperparameters used for EPGAT for each organism are listed in Table~\ref{table:parameters}.

In order to improve the generalization and robustness of our models, we applied the regularization techniques dropout\cite{srivastava2014dropout} and L2 regularization. 
After every GAT layer, we added a dropout layer with probability $p$, specified for each organism as summarized in Table~\ref{table:parameters}. We also added a regularization based on the L2-norm of the model's hyperparameters when computing its gradient. This causes the weights to decay to smaller values, thus biasing the learning procedure towards simpler solutions.

The performance was assessed using the area under the ROC curve (AUC). The ROC curve shows the performance of a classification model in distinguishing between two classes at distinct thresholds settings, by plotting the corresponding True Positive Rate (TPR, in the y-axis) and the False Positive Rate (FPR, in the x-axis). The AUC score may be interpreted as the probability that the model ranks a random positive instance more highly than a random negative instance. Therefore, higher AUC scores indicate better classification models. The choice of this metric aims to allow a qualitative comparison with previous works and to better deal with the inherent class imbalance in our datasets in contrast to measures such as accuracy.

\begin{table}[t]
  \begin{center}
\caption{Hyperparameters used for training the GAT models for each organism.} % title of Table
\label{table:parameters}
\resizebox{0.8\columnwidth}{!}{%
\begin{tabular}{c|c|c|c}%{ |p{2cm}||p{1cm}|p{1cm}|p{2cm}|  }
\toprule
Parameter & \textbf{\yeast} & \textbf{\fly} & \textbf{\coli} \& \textbf{\human} \\
 \midrule
Learning rate & 0.005 & 0.005 & 0.005 \\
Weight decay &  2e-4 & 5e-4 & 5e-4 \\
Hidden units & 12 & 16 & 8 \\
Attention heads & 8 & 8 & 8 \\
Dropout & 0.3 & 0.6 & 0.4 \\
 \bottomrule
\end{tabular}}
  \end{center}
\end{table}

\subsection{Baselines}
\label{S:Baselines}
As baselines in our study, we adopted approaches from the three classes of computational strategies more common in the related literature, namely, network topology measures, shallow ML algorithms, and node embedding coupled with deep learning. 

As for topology measures, we used degree centrality (DC), which calculates the number of neighbors of a node $i$ in the
network; neighborhood centrality (NC), which is based on edge clustering coefficient; and local average connectivity (LAC), which evaluates the
local connectivity of node $i$'s neighbors. For further details about these measures, we refer reader to the survey by Li \etal \cite{survey}.

The traditional, shallow ML algorithms multilayer perceptrons (MLP) and support vector machines (SVM) were also applied as baselines. 
%Also, we benchmark our method against other standard, shallow ML approaches, namely multilayer perceptrons (MLP) and support vector machines (SVM). 
These algorithms do not explicitly handle graph data as input. Therefore, standard feature vectors for each gene were constructed by concatenating their information from our multiomics dataset  (\ie gene expression, orthology, and subcellular organization) and extracted from the STRING PPI network topology (\ie node degree) into a single vector. Next, feature vectors for all genes are concatenated in a single table and used as input for training these models. 
The MLP network was configured with a single hidden layer of 32 units and trained with a dropout rate of $0.2$ to reduce overfitting. %, this configuration is the same both for the MLP with and without the graph embedding input features.
The SVM was trained using the radial basis function (RBF) kernel.

% The input dataset were composed by the tabular data in our multiomics datasets, concatenated together with the node degree extracted from String's PPI network. 
 
Finally, we compared our results against models built upon network features extracted by the node2vec (N2V) embedding method, as applied in previous works (\eg \cite{zeng2019deep, deepep,deephe}). N2V uses biased random walks on graphs to learn a graph embedding for each node, which captures the network structure in a regular format. These embeddings are then used as additional features for the MLP training set, so that we carry out a fair performance comparison with our GAT model.
The parameters used for training the N2V-based MLP model were the same as in the comparison with shallow ML algorithms.

% \section{Experiments and Results}
\section{Results and Discussion}
\label{S:Experiments&Results}
Our experiments were implemented in Python using the sklearn library \cite{scikit-learn} for general training and evaluation methodology, and for the SVM model; Pytorch  \cite{paszke2017automatic} for the MLP models; and Pytorch Geometric extension \cite{fey2019fast} for the graph-based models, namely EPGAT and N2V. %Our implementation is openly available\footnote{https://github.com/JSchapke/essential-gene-detection}.

We setup our experiments aiming at i) comparing our approach to other network-based and ML  approaches (both shallow and deep learning methods) used for gene or protein essentiality prediction and ii) assessing the impact over performance of the different choices of PPI databases and multiomica datasets. Performance for the trained models are shown as the mean and the standard deviation of ten runs using random stratified splits of the dataset. We carried out statistical comparisons by means of two-tailed Student t-test with a 95\% confidence level to evaluate the significance of different metrics.

\subsection{Comparison with network topology-based methods}
\label{S:CompareTopology}

Our first batch of experiments compared the proposed EPGAT model to the network centralities DC, LAC, and NC. For the analysis of topology-based methods, genes in the testing set were ordered by their centrality measure in a descending order and the resulting rank was used for AUC score analysis. 

\begin{figure*}[t]
    \centering
    \includegraphics[width=0.95\textwidth]{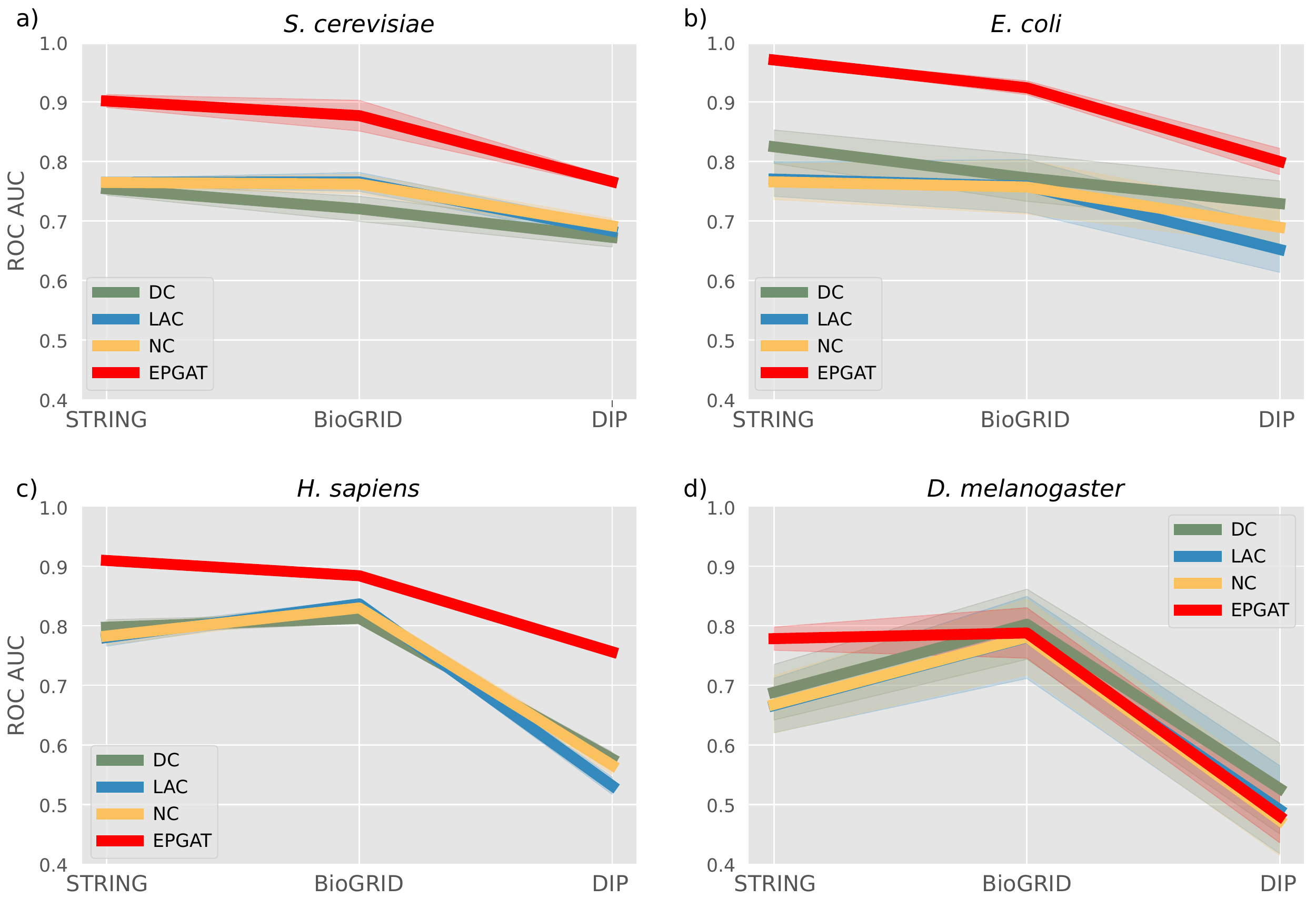}
    \caption{Performance of EPGAT and network-based methods on each network database and organism. Metrics are averaged for 10 repetitions and the shaded areas display standard deviation.}
    \label{auc_network}
\end{figure*}

The results for this comparison considering the three PPI networks, namely STRING, BioGRID, and DIP, are shown in Figure~\ref{auc_network}. In most organisms and networks, our approach had a significant improvement over the baselines. Except for the \fly DIP and BioGrid networks, for which none of the comparisons achieved statistical significance, and for the \coli DIP network, in which EPGAT was only significantly superior than LAC, in all other comparisons EPGAT presented the highest performance with statistical significance ($p < 0.05$).

Surprisingly, the topology-based methods were less consistent than EPGAT. This is mainly perceived in the results for \coli and \fly. 
The improvement brought by the proposed model in terms of performance values and stability is caused by two main factors. First, these baseline methods are analytical approaches that provide only a crude simplification of nodes patterns and roles within the PPI network, thus probably only partially capturing the information correlated to gene essentiality that it is expected to be contained within this type of biological evidence (\ie protein interaction networks). In contrast, GAT is an expressive neural network that is able to model and detect complex relationships by directly analyzing the complete graph structure, as opposed to centrality metrics derived from it. Second, our EPGAT model parses additional datasets, which provides richer and more complete information for the prediction task. As previously discussed, gene or protein essentiality is a multifactorial phenomenon, which is corroborated by our results.
A distillation of these two factors is explored in a data ablation study presented in Section~\ref{S:DataAblation}

Regarding the poor performance of EPGAT for \fly, we note that the networks provided by BioGRID and DIP are severely restricted for this organism. First, the average degree for these networks is relatively low as compared to STRING and to other organisms. Second, while our dataset contains over twelve thousand labeled genes for the \fly, the DIP and BioGRID networks only map interactions among 625 and 1,778 of them, respectively, such that the training set is also limited in contrast to other scenarios. Finally, it should be noted that the \fly essentialome was by far the most challenging one due to its severe class imbalance (1.95\% of labeled genes are classified as essential). Considering the total number of nodes in the networks gathered from BioGRID and DIP databses, only 4.21\% and 3.18\% of them are essential genes. These constrictions, both in the number of samples and skewness of instances, are hard to be overcome by statistical methods. In the BioGRID network, EPGAT achieved comparable results to the baselines, whereas in the DIP network every method was equivalent to random guessing. We note, however, that the \fly STRING network contained over 8,000 labeled genes, and in this case, EPGAT had significant improvement over the baselines, which shows that given enough data our approach can learn important features even in highly biased datasets.

Noticeably, the DIP network had the worst performance on every organism. For the \fly and the \human networks, the performance for most methods were equivalent to random guessing. EPGAT was able to introduce improvements for the Human DIP network, but its performance was still inferior as those for BioGRID and STRING. As it was briefed in Table~\ref{datasets_table}, DIP is the database with the lowest number of mapped interactions. For the \fly and \human organisms, DIP provides extremely sparse networks, with average degree of 1.44 and 1.6, respectively, in contrast to 4.47 for \yeast and 4.18 for \coli. This characteristic impairs classification performance given the underlying functioning of GATs, which depends on a node's connectivity to update its embedding. %  \ie updating a node's embedding based on its neighbors.
Nonetheless, even for \yeast, for which DIP presents the highest average degree and ratio of labeled genes (\ie 89.15\% of nodes in the network) among all organisms, its performance is inferior than denser network choices. %, namely STRING and BioGRID. 
This is an important observation since much work on the field rely solely on the DIP PPI network to evaluate algorithms \cite{ref4, ref6}. In general, BioGRID and STRING had comparable performance among the analytical methods, although our model had a slightly better edge with the STRING network. Therefore, in the next experiments, we use it as a basis for our model.

\begin{figure*}[!t]
    \centering
    \includegraphics[width=0.95\linewidth]{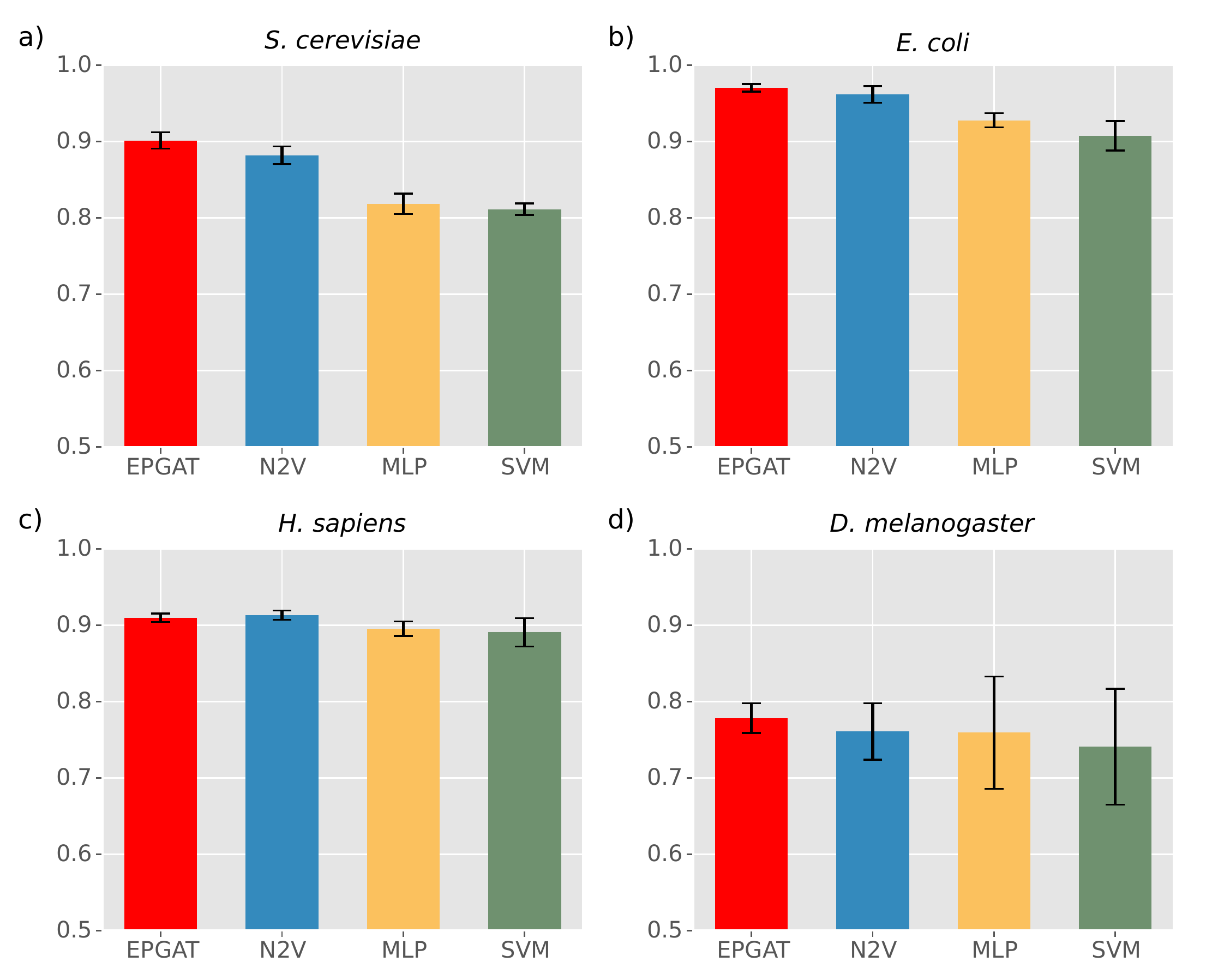}
    \caption{Comparison among EPGAT and other machine learning approaches based on 10 random splits of labeled data. The models were evaluated using interaction information from the STRING network. Black bars display the standard deviation. }
    \label{auc_ml}
\end{figure*}

\subsection{Comparison with machine learning methods}
\label{S:ShallowML}

%Since ML methods are becoming increasingly recurrent in the task of identifying essential genes and proteins, 
We compared the proposed approach with three ML classifiers: MLP, SVM, and a node2vec-based model to extract features that are further analyzed using a MLP (N2V). All classifiers were trained using the collected multiomics dataset (see Section~\ref{S:Data}). Nonetheless, while the standard, shallow ML methods (SVM and MLP) relied on a structured table with features extracted from PPI networks and other omics data, the graph-based methods (\ie EPGAT and N2V) are tailored to the task of learning the gene essentiality-related patterns directly from the input network. We note, however, that their underlying functioning for creating a low-dimensional feature vector for each node in the network is different, as well as the strategy to combine multiomics data with network-based features.

Figure~\ref{auc_ml} displays the performance of the EPGAT model in contrast to the baseline ML methods. All models use as basis the STRING PPI network. Our approach notably improved the prediction performance over MLP and SVM for \yeast, \coli, and \human, achieving statistically significant differences ($p < 0.05$) in these comparisons. MLP was significantly superior than SVM for \coli, but in all other organisms the average performance for both shallow ML methods was very similar.

Regarding the comparison between EPGAT and N2V, overall, we observed a very competitive performance of our model in relation to the node2vec embedding, which is the state-of-the-art method used to learn from graph structured data in related works. EPGAT achieved a superior performance with statistical significance for \coli ($p = 0.033$) and \yeast ($p = 0.001$). For \fly, the difference was not statistically significant ($p = 0.20$), nonetheless, EPGAT had a very positive impact on the stability of the model, notably reducing the standard deviation obtained from the ten random splits of data in relation to the other methods, including N2V. Finally, EPGAT and N2V achieved very close performance for identification of human essential genes (90.95$ \pm $ 0.0054 for EPGAT versus 91.30$ \pm $0.0059 for N2V, with $p = 0.18$).

Thus, in general, our approach improved the identification of essential genes over previous ML methods for all organisms, either by increasing the average AUC score or by generating a more robust model (\ie presenting lower variance on predictive performance). 
Our results show the importance of a more elaborate approach to deal with graphs in this prediction task. 
The advantage in approaches that maximally preserve properties and information from graph whilst identifying gene essentiality patterns is especially clear in the comparison between the N2V and MLP, which are two models based on the MLP learning algorithm. By adding the node embeddings generated with node2vec to its feature set, N2V was significantly better ($p < 0.01$) than the shallow MLP for all organisms except \fly.

\begin{table}[!b]
  \begin{center}
    \caption{Data ablation study shows AUC performance for the GAT model with incrementally more omics data. DC centrality is used as the baseline.}
    \label{table1}
    \resizebox{0.8\columnwidth}{!}{%
    \begin{tabular}{l|c|c|c|c|c|c} 
    \toprule % <-- Toprule here
     \multirow{2}{*}{\textbf{Method}} &  \multicolumn{3}{c|}{\textbf{\yeast}} & \multicolumn{3}{c}{\textbf{\coli}} \\
      {} & $DIP$ & $BioGRID$ & $STRING$ & $DIP$ & $BioGRID$ & $STRING$\\
      \midrule % <-- Midrule here
        DC & 67.22 & 72.01 & 75.38 & 72.90 & 77.21 & 82.44 \\
        GAT & 71.57 & 88.58 & \textbf{90.43} & 71.47 & 89.90 & 96.43 \\
        + Expression & 72.32* & \textbf{89.02} & 90.21& 78.02* & 92.09* & 96.71 \\
        + Sublocalizations & 72.57 & 88.71 & 90.01 & --- & --- & --- \\
        + Orthology & \textbf{76.57}* & 87.66 & 90.13 & \textbf{79.96}* & \textbf{92.35} & \textbf{97.02} \\
    \toprule % <-- Toprule here
     \multirow{2}{*}{} & \multicolumn{3}{c|}{\textbf{\human}} & \multicolumn{3}{c}{\textbf{\fly}} \\
      {} & $DIP$ & $BioGRID$ & $STRING$ & $DIP$ & $BioGRID$ & $STRING$\\
      \midrule % <-- Midrule here
        DC & 57.64 & 81.22 & 79.75 & 52.65 & 80.25 & 68.83 \\
        GAT & 63.70 & 87.81 & 90.34 & \textbf{48.61} & 77.20 & 72.67 \\
        + Expression & 79.63* & 88.04 & 91.28*  & 40.90 & 78.31 & 73.71 \\
        + Sublocalizations & \textbf{80.19} & \textbf{89.09}* & \textbf{91.52} & 48.30 & \textbf{79.87} & 73.95 \\
        + Orthology & 75.61 & 88.39 & 90.95 & 48.01 & 78.78 & \textbf{77.82}* \\
      \bottomrule % <-- Bottomrule here
    \end{tabular}
    }
    \\
    \scriptsize{* Statistically significant ($p<0.05$) performance improvement in comparison to previous model in the incremental analysis.}
    %* performance is the same since sublocalizations were not available. %for \coli.
  \end{center}
\end{table}
\subsection{Data ablation study}
\label{S:DataAblation}

To elucidate the importance of each type of biological evidence in our multiomics-based approach, we carried out an ablation study for the proposed EPGAT model. Performance of EPGAT was evaluated with the PPI network (\ie, STRING, BioGRID, or DIP) without any additional attributes, followed by incremental inclusion of other omics data to investigate their impact over the model AUC score. Results are shown in Table~\ref{table1}, in which the best performance per organism network are highlighted in bold.

We perceive mixed results and a variance in performance depending not only on the choice of data, but on the organism analyzed as well. Nonetheless, in general, additional types of biological evidence seems to positively impact models predictive power as compared to the exclusive use of PPI networks. Notably, gene expression data presented a great potential to boost performance over every choice of PPI network and organism. The only exception was for the \fly DIP model, which had a very unstable performance due to the sparseness of this network (see Figure~\ref{auc_ml}-d) and the large class imbalance in the labeled dataset. Therefore, results for this specific scenario may not be statistically relevant as the previous AUC performance was already inferior to a random classifier.
Interestingly, the use of additional omics data with the DIP network for \yeast, \coli, and \human was able to decrease the performance gap in relation to the other networks, despite the general limitation of sparseness observed in the DIP database.

Subcellular localization information had only a slight quantitative impact on the results, but in four out of the 12 scenarios analyzed (\ie combinations of PPI networks and organisms), the EPGAT model trained with PPI plus gene expression and sublocalization data led to the best performance. This is an interesting result, since although the complete multiomics models were the best in 5 of these scenarios, for \coli we could not assess models trained with sublocalization due to data unavailability. Therefore, it is not possible to assert that models with ortholog were in fact the best framework, or if this result is associated with the lack of sublocalization information for \coli.
%Orthology information was the least helpful of alll datasets, in most cases it either degrades the AUC score or only slightly increases it (though we also note as they were the last dataset added they were already the most likely to cause dimensionality issues). 

Finally, we note that even without other sources of omics data, our EPGAT model surpasses the DC baseline in most cases. A statistically better performance ($p<0.05$) of EPGAT in relation to DC was observed in all models, except for the \fly's DIP and BioGRID networks and for the \coli's DIP network. In these two models, DC achieved a higher AUC score, although without statistical significance considering a 95\% confidence level.

\subsection{Analysis of PPI thresholds on STRING network}
\label{S:PPIThreshold}
%\section{Discussion}
%\label{S:Discussion}

The impact on model performance according to the choice of PPI network database is quite noticeable from our experimental results summarized in Table~\ref{table1}. This observation led us to examine the impact of filtering out the networks used as input for EPGAT based on distinct confidence measures, as in the previous experiments we kept only interactions with a STRING confidence score higher than 0.5.

\begin{figure}[b]
     \label{fig:stringScore}
    \centering
    \includegraphics[width=\linewidth]{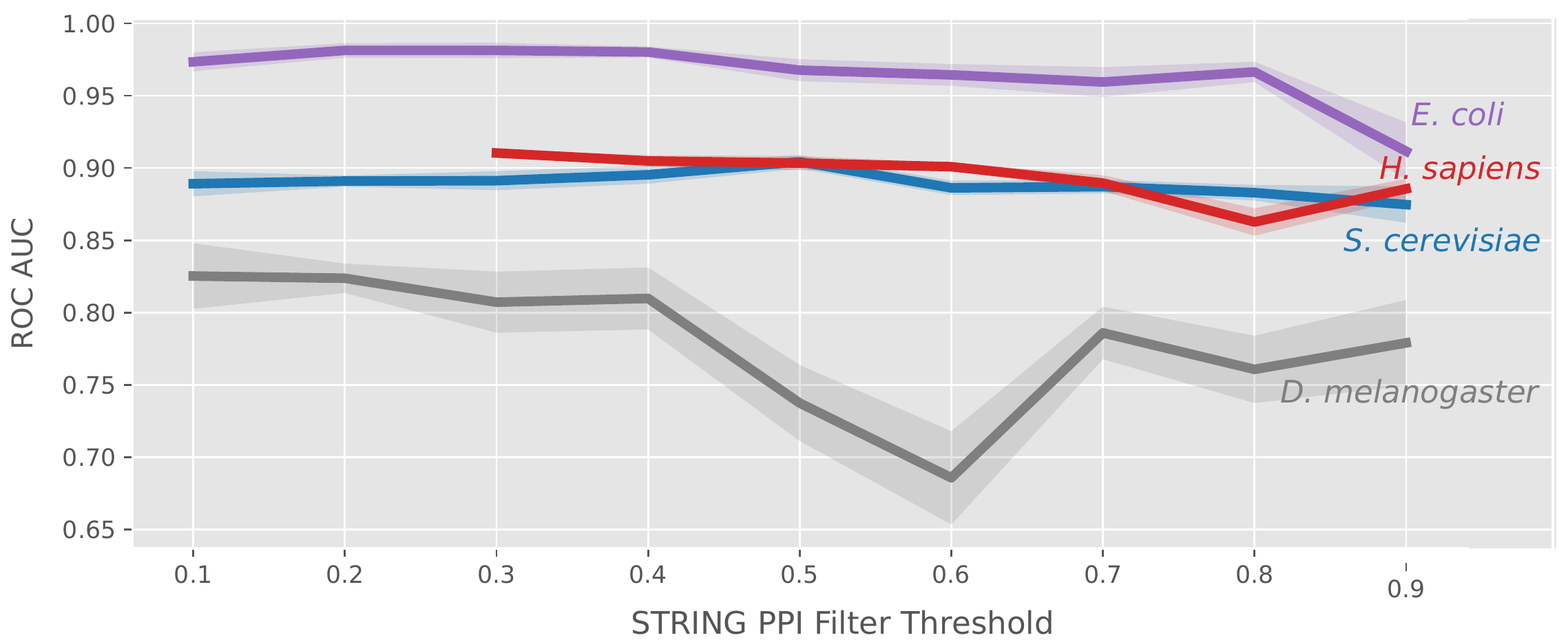}
    \caption{Performance with the STRING network with different filtering thresholds Results on human were truncated due to memory limitations.}
\end{figure}

We rerun the experiments with the EPGAT model in all organisms using different filtering thresholds. Results are given in Figure~\ref{fig:stringScore}, in which the confidence thresholds are indicated in the x-axis. Experiments for \human were truncated at a confidence score of 0.3 due to memory constraints with denser networks. 

The best AUC score was obtained at a threshold of 0.2 for \coli (98.13\%), of 0.3 for \human (91.00\%), of 0.5 for \yeast (90.42\%), and of 0.1 for \fly (82.53\%). Hence, there is a clear trend towards smaller thresholds achieving better performance. This finding corroborates with our previous assertions on dense networks carrying more information about gene essentiality, even if at the cost of more misinformation, and yielding better results for our approach. Nonetheless, a surprising observation is that for most organisms, EPGAT performance was relatively consistent even with largely diverging threshold values and, thus, different number of interactions within the PPI network.  The variation between the best and the worst performance was relatively small for \coli (0.069), \yeast (0.029), and \human (0.047). This result reinforces the robustness and potential of EPGAT for this prediction task.

For the \fly dataset, we observed an erratic behavior. Not only the average performance was unstable across distinct PPI filter thresholds, but also the standard deviation was consistently higher when compared to other organisms. We believe this is caused by the limited volume and class imbalance characteristics for the essential genes data. % related to \fly. 
\fly dataset comprised only 161 positive labeled genes in the STRING PPI network used in our work, which causes any perturbation, even if a small one, to affect classification and exert a significant impact on the AUC score. Moreover, as we may see in Table~\ref{datasets_table}, the \fly STRING PPI network has a relatively large dimension when compared to \yeast and \coli, which aggravates the situation as changes in the filtering thresholds lead to a more prominent perturbation on the structure of the network. Finally, we note that EPGAT models could be enhanced by optimizing the interactions filtering threshold in a case-by-case basis, which was not explored in the current work.

\subsection{GAT model interpretation}
\label{S:GATInterpetati}
Our model is based on an attention mechanism, which weights every interaction among two genes $i$ and $j$. 
Intuitively, the weight is a value $\alpha_{i, j} \in [0, 1]$ that assigns the importance of node $j$ for the prediction returned for node $i$.
In other words, the weights can be seen as the influence that a gene $j$ causes in the classification of another gene $i$. Therefore, learned attentional weights may provide benefits in model interpretability. 

\begin{figure}
    \centering
    \includegraphics[width=0.9\linewidth]{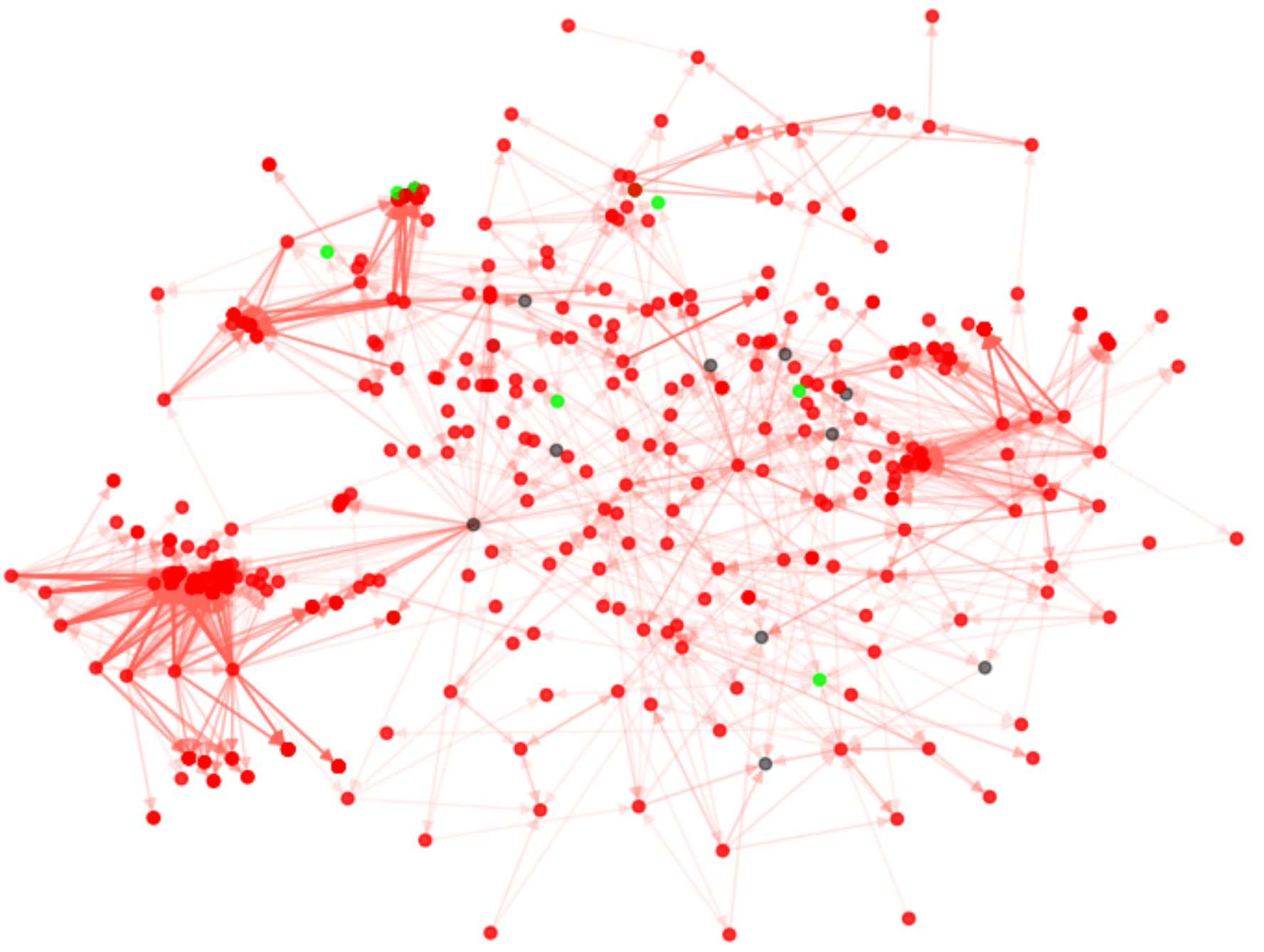}
    \caption{\textbf{Subgraph of the human network with the STRING database.}  The shade of each edge in the network represent the weight given by GAT's attention mechanism, darker edges correspond to interactions which the model assigned more weight for predictions, the opposite for lighter. Green and red nodes correspond to essential and not essential genes respectively, while black nodes to genes without a label. }
    \label{fig:attention}
\end{figure}

We visualize the attention weights for the \human STRING PPI network in Figure~\ref{fig:attention}. For better visualization purpose, we obtained a subgraph from highly connected nodes to reduce network dimension.
The weight is displayed by the transparency of each edge, such that strongly shaded edges means that the interaction had a greater influence on the classification of the target gene. Green and red nodes correspond to essential and non-essential genes, respectively, while black nodes refer to network genes without a label (\ie not comprised in the collected essential genes dataset).

Noticeably, we see an interesting phenomenon. The nodes with very few connections do not affect nor are strongly affected by their neighbors, which means that the model relies on the features given by the additional datasets for their prediction.
Highly connected nodes (\ie hubs), on the other hand, show the opposite effect as they greatly affect the predictions of their neighbors. Complementarily to the \textit{centrality-lethality} rule that affirms that hubs have a higher likelihood of being essential, our model indicates that genes interacting with hubs are likely to have similar essentiality status. In this network visualization, this is characterized by the highly interacting negative labeled genes in the lower left of Figure~\ref{fig:attention}. 

\section{Conclusion}
\label{sec:conclusion}
%Despite the importance in the identification of essential genes and proteins due to their significance for biology and pathology, this is still an open problem in genomics and bioinformatics. %Several challenges exists, but the evolving concept of essentiality patterns and the inherent noise in genomics data are some of them. In this study, we hypothesized that the integration of a deep learning method able to fully explore the graph-based data and additional evidence provided by multiomics data would improve precision in the prediction of essential genes and proteins.
 %we formulated the task of gene essentiality prediction as a task of node classification by means of GNNs, and proposed an approach based on GATs, a specific type of GNNs. Our approach is able to fully explore the graph-based data provided by PPI networks in  and additional evidence provided by multiomics.
In this study, we approached the gene essentiality prediction problem by using powerful GATs, a type of GNNs, combined with the integration of multiomics datasets. While most of the previous research on the field evaluate their algorithms on mostly one or two organisms (especially model organisms), we adopted a more comprehensive benchmark that included \human datasets. We showed that our approach, named EPGAT, achieved a significant improvement in performance compared to other network-based and machine learning methods commonly used in the field. % and competitive performance against state-of-the-art methods such as node2vec embedding method% other GNNs.  
Notably, results obtained by EPGAT were at least comparable (when not superior) to node2vec, a state-of-the-art method for gene or protein essentiality prediction (used, for instance, in  \cite{zeng2019deep, deepep,deephe}). While maintaining a slight edge, EPGAT still has the benefits of having a more straightforward training procedure and shorter training time. 

Our experiments indicated that denser PPI networks are more informative and reliable for gene essentiality prediction. However, even under the challenged posed by network sparseness (\ie \fly network in our study), EPGAT was the most robust ML method.  Lastly, we observed that our approach can also be helpful for insights over gene relations, as the GATs' attention mechanisms is a measure of the influence among interacting genes. Further investigation towards interpretability of model attentional weights may help shed light on network-based features of genes and proteins essentiality patterns. Other interesting research directions include (i) expanding our approach to capture the context-dependent nature of gene essentiality, for instance, by integrating more omics datasets, and (ii) incorporating node-level feature selection techniques in our model to control dimensionality and improve its generalization power.

%Finally, we note that results obtained by GAT are at least comparable (when not superior) to node2vec, a state-of-the-art method for gene or protein essentiality prediction (used, for instance, in  \cite{zeng2019deep, deepep,deephe}). While maintaining a slight edge, GAT still has the benefits of having a more straightforward training procedure and shorter training time.
%Finally, 

%To the best of our knowledge, our work is the first to explore  graph-based deep learning in this context and, as we show how competitive this algorithm is in contrast to state-of-the-art approaches such as node2vec embedding method.

% Moving forward we suggest fellow researches to evaluate their models with denser networks such as STRING and BioGRID, which should provide a better, more consistent and reliable performance.

% Future work. 
% Although throughout this study we considered gene essentiality in a binary classification problem, it's most well interpreted as a difuse property which is conditioned 

%\appendices
%\section{}

\ifCLASSOPTIONcompsoc
  \section*{Acknowledgments}
\else
  \section*{Acknowledgment}
\fi

This work was partially supported by CNPq/Brazil and by CAPES Finance Code 001.

\ifCLASSOPTIONcaptionsoff
  \newpage
\fi




\bibliographystyle{IEEEtran}
\bibliography{references}
\vfill

%\vfill


\end{document}